\documentclass[
article,
 amsmath,
 amssymb,
 aps,
 twocolumn,
prl,
superscriptaddress,
nofootinbib
]{revtex4-2}

\usepackage{graphicx}
\usepackage{dcolumn}
\usepackage{bm}
\usepackage{color}
\usepackage{tikz}
\usetikzlibrary{arrows.meta,positioning}

\begin{document}
\title{
A Dynamical Mechanism for Irreversibility in Cyclically Driven Amorphous Solids
}
\author{Sauvik Chatterjee}
\affiliation{Department of Solar Energy and Environmental Physics, Jacob Blaustein Institutes for Desert Research, Ben-Gurion University of the Negev, Sede Boqer Campus 84990, Israel}
\author{Asaf Szulc}
\affiliation{Department of Physics, Ben Gurion University of the Negev, Beer Sheva 84105, Israel}
\author{Ido Regev}
\email[E-mail: ]{regevid@bgu.ac.il}
\affiliation{Department of Solar Energy and Environmental Physics, Jacob Blaustein Institutes for Desert Research, Ben-Gurion University of the Negev, Sede Boqer Campus 84990, Israel}
\affiliation{Department of Physics, Ben Gurion University of the Negev, Beer Sheva 84105, Israel}

\date{\today}

\begin{abstract}
Amorphous solids subjected to athermal quasistatic oscillatory shear undergo a transition from periodic reversible dynamics to irreversible diffusive dynamics at yielding. How irreversibility arises in such deterministic, strongly dissipative systems remains unclear. Here we directly test the proposal that post-yield irreversibility originates from chaotic dynamics and exponential sensitivity to initial conditions. Contrary to this interpretation, perturbations initially contract rather than grow, even in the irreversible regime, with nearby trajectories remaining close for extended periods. Separation occurs only through rare branching events, after which the distance grows diffusively at the rate expected for independent trajectories. The waiting times to branching are exponentially distributed, defining a steady-state branching rate that vanishes in trained reversible limit cycles and becomes finite above yielding. A mean-field soft-spot model reproduces these branching statistics and reveals their microscopic origin: a small perturbation can reverse the activation order of two nearly degenerate plastic instabilities, altering the subsequent sequence of plastic events. These results show how local stability and global irreversibility can coexist in a dissipative many-body system and identify instability-selection-induced branching as a distinct dynamical route to irreversibility in driven amorphous solids.
\end{abstract}

\maketitle
%
%
%
%
%
%
%
\section{Introduction}
Amorphous solids are ubiquitous in nature and in industrial applications. 
A striking feature of these materials is that when subject to oscillatory shear, they exhibit a transition between two qualitatively different dynamical regimes at the yielding amplitude. At sufficiently small strain amplitudes they reach a periodic limit cycle in which the same sequence of plastic events repeats from cycle to cycle \cite{regev2013onset,hexner2020periodic,fiocco2014encoding}. Above a critical strain amplitude, however, the dynamics is irreversible
and the particle trajectories become diffusive. This is known as the reversibility-irreversibility transition in amorphous solids \cite{fiocco2013oscillatory,regev2013onset,keim2013yielding,tjhung2016criticality,schreck2013particle,reichhardt2023reversible,priezjev2016reversible,mungan2025self}. Similar reversibility-irreversibility transitions have been observed in colloidal suspensions, granular materials, emulsions, superconducting vortices, and other driven disordered systems \cite{corte2008random,pine2005chaos,royer2015precisely,mobius2014ir,Granular1,weijs2015emergent,jeanneret2014geometrically,reichhardt2023reversible,suda2025yielding,mangan2008reversible,xu2013contact,shohat2022memory}, suggesting that they are a generic feature of cyclically driven many-body systems. In amorphous solids, the reversibility-irreversibility transition is closely related to yielding under oscillatory deformation \cite{parley2022mean,galloway2022relationships,bonn2017yield} and is often viewed as its dynamical manifestation. Consequently, it has been studied extensively in experiments \cite{keim2013yielding,nagamanasa2014experimental}, particle-based simulations \cite{yeh2020glass,das2020unified,leishangthem2017yielding,schreck2013particle}, elastoplastic models \cite{kumar2022mapping,szulc2024overlapping}, and more abstract glass models \cite{suda2025yielding}.

One of the central open questions with regard to the transition is the manner in which such systems transition from a mechanically reversible regime, in which the dynamics converges to limit cycles, to an irreversible and diffusive regime as the strain amplitude is increased.
Several attempts have been made to understand this transition within the framework of non-equilibrium statistical mechanics. Proposed interpretations include transitions into absorbing-state \cite{ness2020absorbing,corte2008random}  and first-order transition scenarios \cite{kawasaki2016macroscopic} associated with metastability \cite{mungan2021metastability}. While these approaches provide important insight into the collective behavior near the transition, they do not directly address the dynamical mechanism through which irreversibility emerges.

From a dynamical-systems perspective, the transition is particularly puzzling. Both athermal quasistatic (AQS) simulations and slowly driven experiments on colloidal and granular systems are strongly dissipative. Unlike inertial molecular dynamics, AQS dynamics consists of a sequence of strain increments followed by energy-minimization steps. Since these minimizations tend to contract perturbations between nearby configurations, it is not obvious how irreversible dynamics can emerge. One would therefore expect the dynamics to converge to an attractor, which under periodic forcing could be either a limit cycle or a chaotic attractor. Indeed, simulations accompanying the pioneering experiments of Pine and coworkers on dilute colloidal suspensions reported positive Lyapunov exponents in the irreversible regime, motivating the idea that irreversibility may originate from chaotic dynamics and exponential sensitivity to initial conditions \cite{pine2005chaos}. Similarly, in earlier work on amorphous solids one of us suggested, on the basis of time-series analysis \cite{kantz1997nonlinear}, that the post-yield dynamics might be chaotic, providing a route to diffusion through a positive Lyapunov exponent \cite{regev2013onset}.

Here we revisit this question by directly studying the response of state-space trajectories to small random perturbations of their initial conditions. This approach is similar in spirit to that used by Pine \textit{et al.}~\cite{pine2005chaos} to estimate Lyapunov exponents. It is also related to the perturbative approach employed by Jaiswal \textit{et al.}~\cite{jaiswal2016mechanical} to study yielding under monotonic deformation as the recovery of ergodicity, by analogy with phase transitions in spin glasses.

Contrary to the conventional chaotic interpretation, we find that perturbations contract rather than grow exponentially, even in the post-yield regime. Nearby trajectories instead remain close for extended periods before separating through discrete branching events, after which their separation grows diffusively. The waiting times before branching are exponentially distributed, defining a branching rate that vanishes for stable trained limit cycles and becomes finite above yielding. A mean-field model of interacting hysterons representing soft spots reproduces these statistics and reveals an instability-selection mechanism: a small perturbation can change which of two nearly degenerate plastic instabilities is activated first, thereby altering the subsequent sequence of plastic events. These results identify trajectory branching as a dynamical route by which local stability and global irreversibility coexist in cyclically driven amorphous solids.
\section{Results}
As discussed above, it was previously suggested, based on time-series analysis of the potential energy of simulations, that irreversibility in the post-yield regime may arise from chaotic dynamics exhibiting exponential sensitivity to initial conditions \cite{regev2013onset}. To directly test for sensitivity to initial conditions, we study the stability of nearby trajectories in a particle-based Kob-Andersen model \cite{bruning2009glass} of a binary, two-dimensional amorphous solid with $N=1000$ particles, subjected to athermal quasistatic shear (AQS).
In the AQS protocol, the system is sheared in small strain increments $\delta\gamma$, each followed by minimization of the potential energy  \cite{maloney2006amorphous}. Cyclic deformation is imposed through the strain protocol
\begin{equation}
0 \rightarrow \gamma_{\rm max}
\rightarrow -\gamma_{\rm max}
\rightarrow 0,
\end{equation}
where $\gamma_{\rm max}$ is the strain amplitude. For this model, yielding is found to be $\gamma_c = 0.065$  \cite{chatterjee2026memory} (see additional simulation details in the Supplementary Material).
\begin{figure}[htbp]
    \centering
    \includegraphics[width=0.52\textwidth]{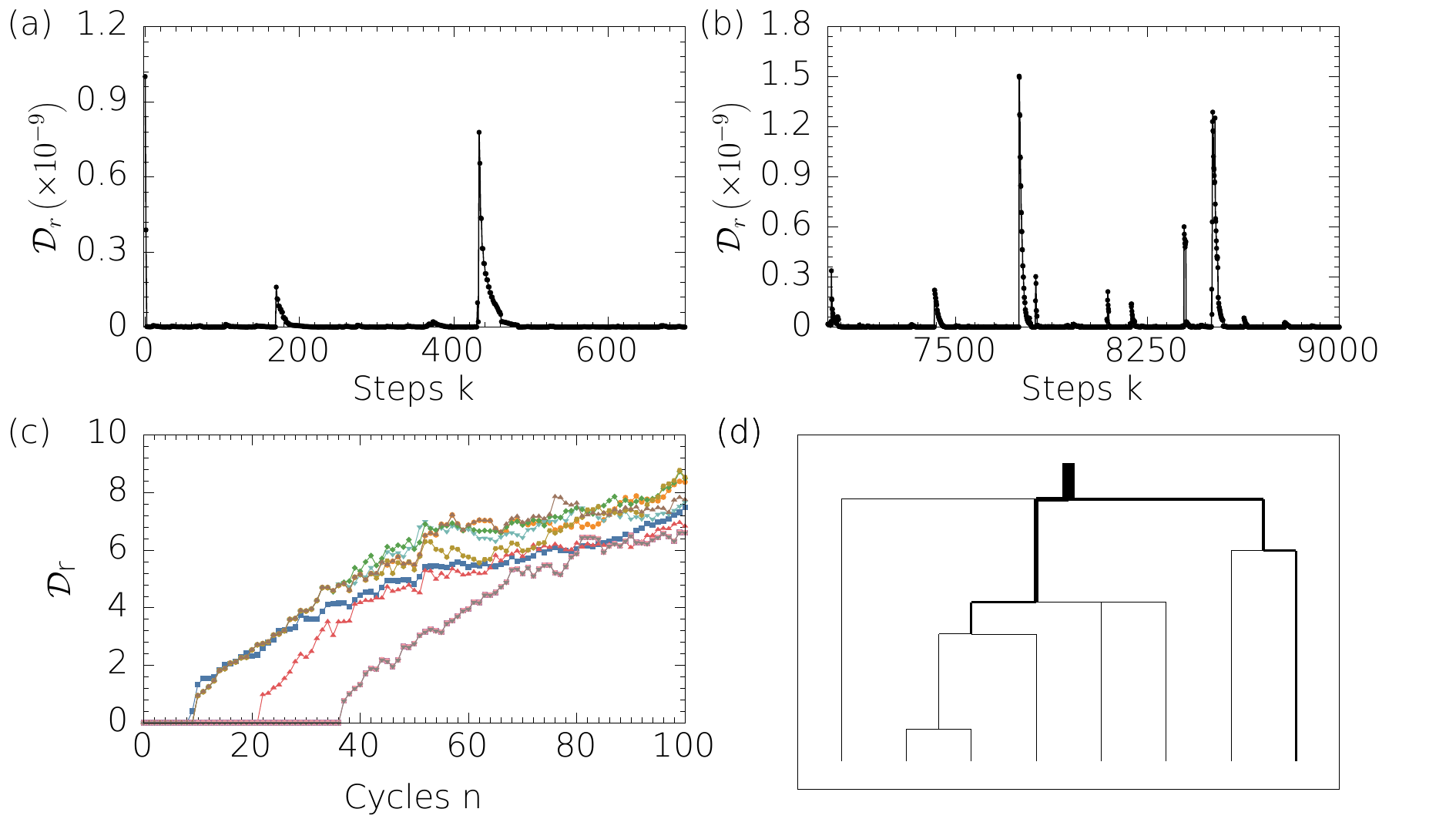} 
\caption{
(a) RMS particle-wise separation $\mathcal{D}_r$ between a trajectory on a limit cycle with $\gamma_{\rm max}=0.06$ and a perturbed trajectory, shown as a function of the number of AQS strain steps $k$ during a forcing cycle. The vertical axis is scaled by $10^{-9}$. We can see that the initial perturbation decays and that small deviations that subsequently emerge also decay exponentially, demonstrating the stability of the limit cycle. 
(b) Same as panel (a), but for a post-yield trajectory with $\gamma_{\rm max}=0.136$. Although the trajectory never becomes periodic, small perturbations decay exponentially, demonstrating that the trajectory is locally stable between separation events. 
(c) RMS particle-wise separation measured at zero strain after $n$ complete forcing cycles for 10 independent random perturbations of the same reference configuration subject to a post-yield $\gamma_{\rm max}=0.136$. Several perturbed trajectories branch from the reference trajectory during the same cycle and subsequently branch from one another. 
(d) Branching diagram corresponding to panel (c). Line thickness indicates the number of perturbed trajectories following the same path at a given cycle.
}
\label{fig1}
\end{figure}

To probe trajectory stability, we apply small perturbations to the initial particle coordinates:
\begin{eqnarray}
r_{x,i}(0) &\rightarrow& r_{x,i}(0) + \delta \cos\theta_i,\\
r_{y,i}(0) &\rightarrow& r_{y,i}(0) + \delta \sin\theta_i,
\end{eqnarray}
where $\delta=10^{-9}$ in LJ units and $\theta_i$ is a uniformly distributed random variable \footnote{We have also checked another form of the perturbation vectors, and found qualitatively similar results.}. We then evolve both the original, reference, configuration and the perturbed systems using the same AQS protocol and monitor their post-yield root-mean square (RMS) particle-wise separation,
\begin{equation}
\mathcal{D}_r(n)=\frac{1}{\sqrt{N}}\|{\bf r}^{(p)}(n)-{\bf r}^{(r)}(n)\|.
\end{equation}
Here ${\bf r}^{(r)}=\{r_{x,1}^{(r)},r_{y,1}^{(r)}\ldots,r_{x,N}^{(r)},r_{y,N}^{(r)}\}$ is the reference coordinate vector, ${\bf r}^{(p)}$ is the corresponding perturbed coordinate vector, and $n$ is the number of forcing cycles. 
For a stable trajectory, such as a limit cycle, we expect perturbations to remain bounded and eventually decay \cite{ott2002chaos}. Indeed, in the limit-cycle regime we observe that small deviations from the reference trajectory decay exponentially fast (Fig.~\ref{fig1}a), indicating exponential stability.
\begin{figure}[htbp]
\centering
\includegraphics[width=0.5\textwidth]{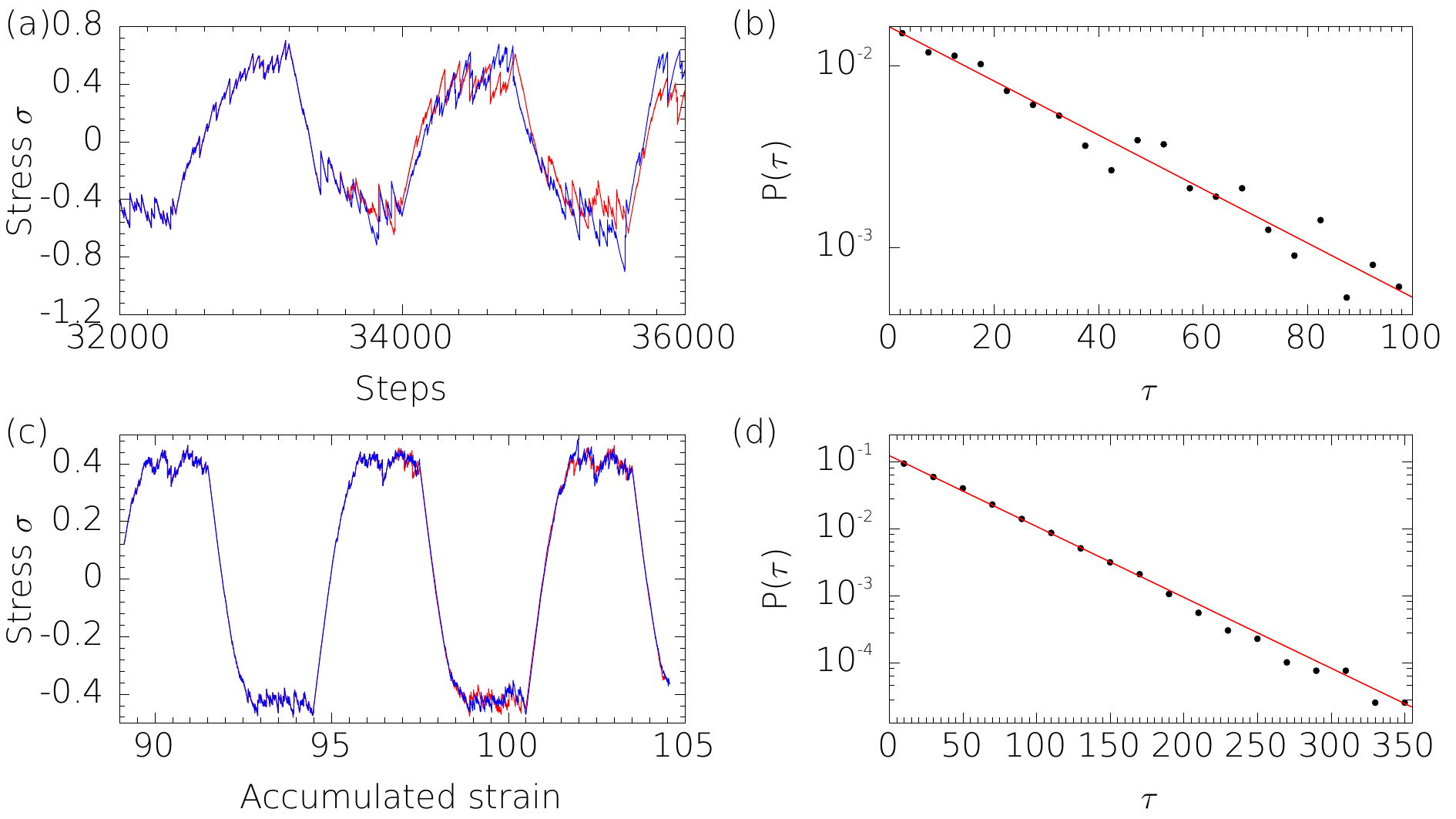} 
\caption{
(a) Trajectory separation in the particle-based simulations subject to post-yield strain amplitude $\gamma_{\rm max} = 0.136$, as reflected in the total stress. The stress in the reference trajectory is shown in blue while the perturbed trajectory is shown in red. Up to a certain simulation step, the stress in the perturbed system is identical to the stress in the reference systems, but the two show clear separation starting from the $\sim 33500$ step.
(b) Waiting-time (in number of cycles $\tau$) in the particle-based simulations in a post-yield steady-state with $\gamma_{\rm max} = 0.136$ exhibiting an exponential distribution.
(c) Trajectory separation in the soft-spot model subject to post-yield strain amplitude $\gamma_{\rm max} = 1.5$, as reflected in the total stress, where colors represent the same as in (a).
(d) Waiting-time (in number of cycles $\tau$) in the soft spot model calculated at a post-yield steady-state with $\gamma_{\rm max} = 1.2$ exhibiting exponential distribution.
The straight lines in panels (b) and (d) are linear fits.
}
\label{fig2}
\end{figure}

Surprisingly, we observe similar behavior in the post-yield regime (Fig.~\ref{fig1}b and supplementary material). Although the post-yield trajectories never reach a limit cycle, nearby trajectories do not separate exponentially fast as expected for chaotic dynamics \cite{ott2002chaos}. Instead, perturbations initially decay and the dynamics remains locally stable. It is therefore clear that conventional sensitivity to initial conditions is not the source of irreversibility in this system, and that the origin of irreversibility must be sought elsewhere.

Although the post-yield dynamics is locally stable, nearby trajectories do not remain close-by indefinitely. Instead, after a seemingly random number of forcing cycles following the initial perturbation, trajectories undergo a sudden separation event at a specific strain step (Fig.~\ref{fig1}c). When we look at only zero strain configurations, we can see that it could take many cycles for a perturbed trajectory to branch from the reference trajectory (Fig.~\ref{fig1}c). 
Furthermore, when several independent perturbations are applied to the same reference trajectory, several of these trajectories often separate from the reference trajectory during the same cycle. (we can see an example for such separation in Fig.~\ref{fig1}c). At later times, some of these trajectories themselves split into distinct branches (Fig.~\ref{fig1}c). The resulting hierarchy of trajectory splits for the example in Fig.~\ref{fig1}c is summarized in the branching diagram shown in Fig.~\ref{fig1}d, where each line represents a distinct trajectory and the line thickness represents the number of perturbed trajectories following the same trajectory. 

We next studied the distribution of waiting times $\tau$ in the post-yield regime, measured as the number of cycles preceding a branching event. We note that separation does not necessarily occur at zero strain as can be seen in Fig.~\ref{fig2}a, but here we study the statistics for complete cycles. 
To identify sustained trajectory branching, we compare configurations at zero strain after each complete forcing cycle. Before branching, the reference and perturbed configurations return to practically identical zero-strain states, whereas after branching their separation is much larger. We define the branching time as the first cycle for which the zero-strain separation exceeds $\mathcal{D}_r=10^{-4}$. 
Fig.~\ref{fig2}b shows the waiting times distribution for 500 reference configurations, each perturbed 10 times after the system reached a steady-state at $\gamma_{\rm max} = 0.136$. As can be seen in the figure, the waiting times follow an exponential distribution. 

To gain further insight into the origin of this behavior, we performed simulations using a model of hysteretic soft-spots with mean-field interactions which was previously shown to exhibit an irreversibility transition at $\gamma_c = 0.97$ \cite{szulc2024overlapping}. In this model, the elementary degrees of freedom are hysteretic elements that behave similarly to soft spots in amorphous solids \cite{manning2011vibrational,mungan2019networks,STZ}. Each soft spot occupies a site, which is subject to a local stress. Hysteretic and multiwell behavior emerges when new soft spots are added at the same site due to driving, and the new soft spots overlap and block previous ones. To probe trajectory stability, we apply small, random, perturbations to the stress field of a zero strain configuration obtained after the system reached a steady-state (see Supplementary Material for details).
In Fig.~\ref{fig2}c, we can see that the soft-spot model (with N=250 sites) undergoes a separation of state-space trajectories in a manner that closely resembles the stress separation in the particle-based model (Fig.~\ref{fig2}a).
In Fig.~\ref{fig2}d we show the waiting time distribution from 1000 reference configurations with $\gamma_{\rm max}=1.2$, each perturbed 10 times, after the system reached a steady-state. We can see that the waiting times in the soft-spot model are also exponentially distributed. The ability of a coarse-grained soft-spot model with mean-field interactions to reproduce this behavior provides strong evidence that the underlying mechanism can be understood at the level of interacting plastic instabilities rather than microscopic particle motion.

\begin{figure}[htbp]
\centering
\includegraphics[width=\linewidth]{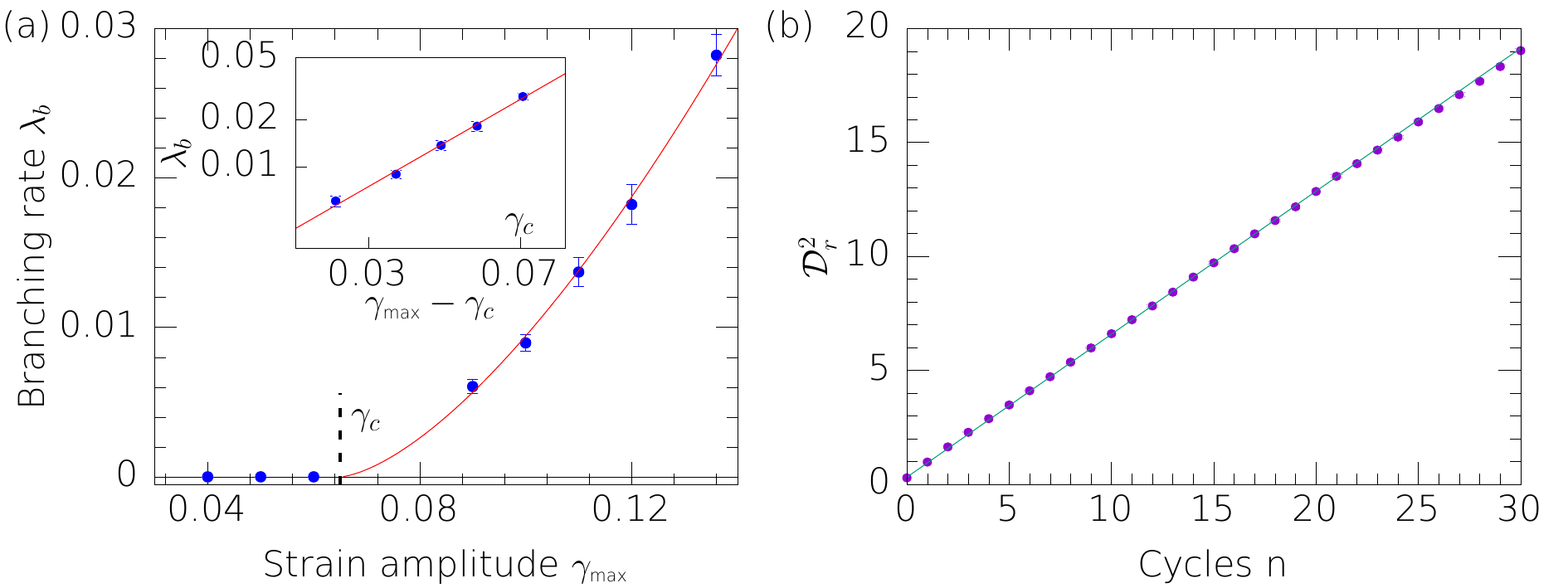}
\caption{
(a) Average steady-state branching rate for different strain amplitudes $\gamma_{\rm max}$ with a power-law fit. 
Inset: post yield values on a log-log scale demonstrating the post-yield power law behavior.  Full details of the averaging procedure are provided in the Supplementary Methods.
(b) Averaged mean-square particle-wise separation in the particle based simulations at a post-yield $\gamma_{\rm max} = 0.136$ steady-state, exhibiting clear diffusive separation (line is a linear fit). Error bars are too small to be visible. 
}
\label{fig3}
\end{figure}

The exponential distribution suggests that, at a coarse-grained level, trajectory branching can be described as a stochastic rare-event process characterized by a branching rate $\lambda_b$ that is constant under steady-state conditions. For sub-yield $\gamma_{\rm max}$, the steady state is a limit cycle, for which we find $\lambda_b = 0$, while the post-yield steady state exhibits a finite branching rate. Over the range of amplitudes studied, $\lambda_b$ increases approximately as a power law $\lambda_b \sim (\gamma_{max} - \gamma_c)^{\beta}$ with exponent $\beta = 1.52(6)$ (Fig.~\ref{fig3}a and inset, where we show the fit used to obtain the exponent). 

A simple interpretation of the exponential waiting-time distribution follows from the approximately exponential distribution of strain intervals between successive plastic events observed under steady-state AQS loading \cite{maloney2006amorphous, tyukodi2019avalanches}. If each plastic event triggers branching with a constant probability $p_b$, the waiting-time distribution for branching is exponential with rate $\lambda_b = p_b\lambda_{\text{pl}}$, where $\lambda_{\text{pl}}$ is the plastic-event rate:
\begin{equation}
f_T(t) = \lambda_b e^{-\lambda_b t}\,.
\label{eq:Dr}
\end{equation}
A detailed derivation is provided in the Supplementary Material (see also \cite{feller1958introduction} for similar derivations).
Since the maximal strain amplitude $\gamma_{\rm max}$ is fixed, the number of AQS strain increments per cycle is constant, and therefore measuring waiting times in cycles or in simulation steps differs only by a constant rescaling of time. The above result is therefore consistent with the exponential waiting-time distributions observed in both the particle simulations and the soft-spot model. 

Once sustained separation occurs, trajectories separate diffusively rather than exponentially. This can be seen in Fig.~\ref{fig3}b, where we show the mean-square particle-wise separation between trajectories, $\mathcal{D}_r^2$ for a steady-state at $\gamma_{\rm max} = 0.136$, averaged over 100 reference trajectories with 10 perturbed trajectories for each reference. The data exhibit diffusive scaling, 
 \begin{equation} 
 \mathcal{D}_r^2(n)\sim n, 
 \end{equation} 
 where $n$ is the number of forcing cycles elapsed since branching.

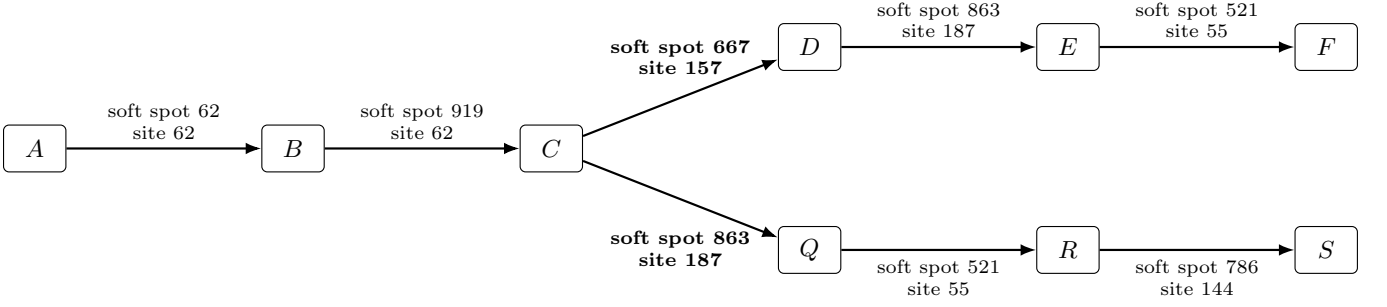
\begin{figure*}[htbp]
\begin{center}
\resizebox{\textwidth}{!}{%
\begin{tikzpicture}[
  state/.style={draw,rectangle,rounded corners=2pt,minimum width=8mm,minimum height=6mm},
  every edge/.style={draw,-{Latex[length=2mm]},thick},
  label/.style={font=\scriptsize,align=center}
]

\node[state] (P1) {$A$};
\node[state, right=25mm of P1] (P2) {$B$};
\node[state, right=25mm of P2] (A) {$C$};
\node[state, right=25mm of A, yshift=13mm] (B) {$D$};
\node[state, right=25mm of B] (C) {$E$};
\node[state, right=25mm of C] (D) {$F$};
\node[state, right=25mm of A, yshift=-13mm] (Q) {$Q$};
\node[state, right=25mm of Q] (R) {$R$};
\node[state, right=25mm of R] (S) {$S$};

\path (P1) edge node[label,above] {soft spot 62\\site 62} (P2)
      (P2) edge node[label,above] {soft spot 919\\site 62} (A);
\path (A) edge node[label,above = 5pt] {{\bf soft spot 667}\\{\bf site 157}} (B)
      (B) edge node[label,above] {soft spot 863\\site 187} (C)
      (C) edge node[label,above] {soft spot 521\\site 55} (D);
\path (A) edge node[label,below = 8pt] {{\bf soft spot 863}\\{\bf site 187}} (Q)
      (Q) edge node[label,below] {soft spot 521\\site 55} (R)
      (R) edge node[label,below] {soft spot 786\\site 144} (S);
\end{tikzpicture}%
}
\end{center}
\caption{
An example of the onset of trajectory branching in the soft-spot model. The branching event occurs inside an avalanche, during the increasing-strain branch of a forcing cycle with $\gamma_{\rm max}=1.2$. The reference and perturbed trajectories follow the same sequence of configurations up to state $C$. At state $C$, the small perturbation changes which of two competing instabilities is selected. In the reference trajectory, soft spot 667 at site 157 is activated, whereas in the perturbed trajectory, soft spot 863 at site 187 is activated. 
}
\label{fig4}
\end{figure*}
\begin{figure}[htbp]
\centering
\includegraphics[width=\linewidth]{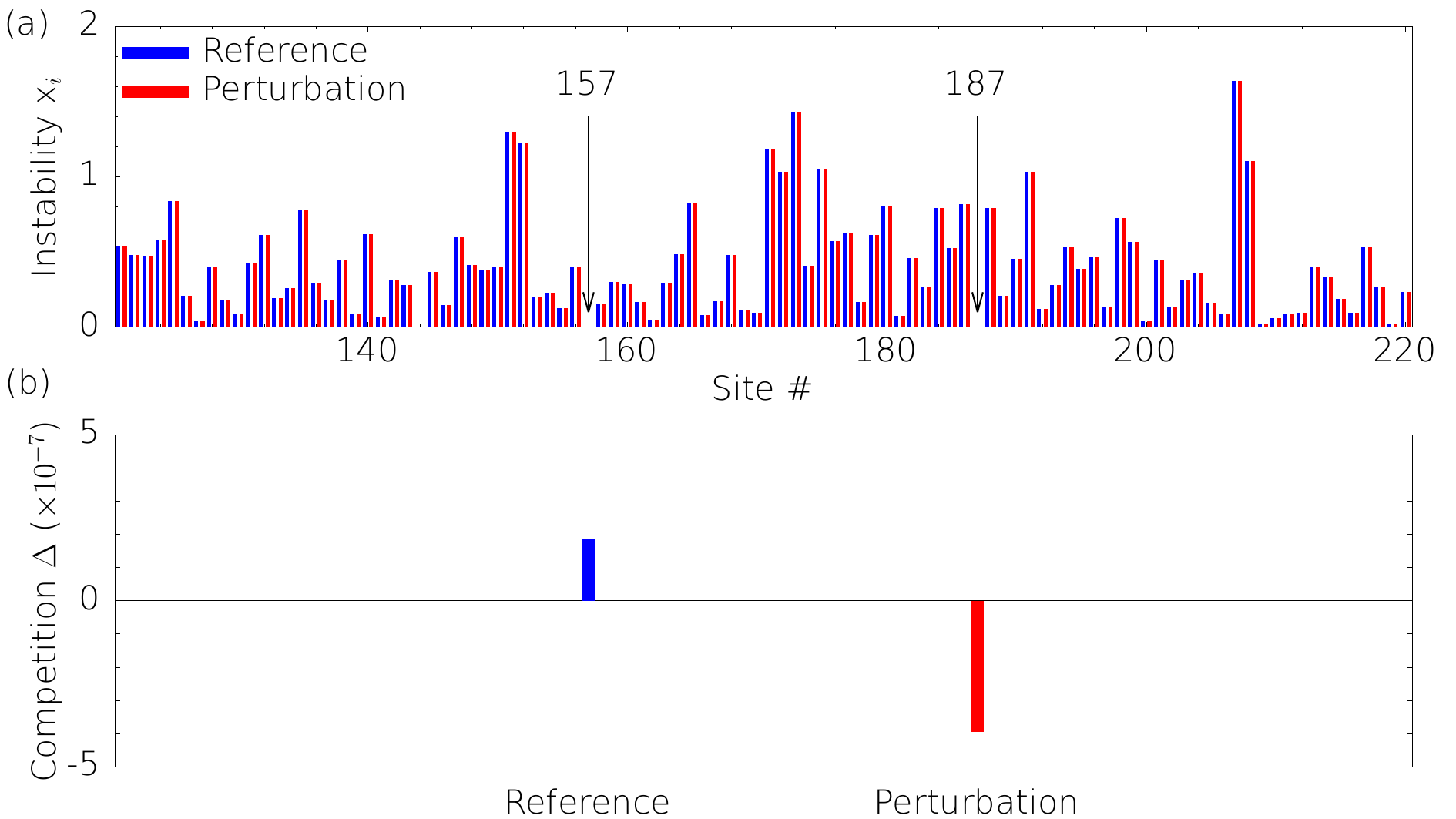}
\caption{
(a) Distances to instability $x_i=\sigma_i^+-\sigma_i$ for state $C$, immediately before the branching event. Sites 157 and 187, marked by arrows, are the only unstable sites with $x_i<0$, and form a nearly degenerate pair of competing instabilities. Because their difference is too small to resolve on the scale of panel (a), their ordering is quantified by the competition variable shown in panel (b). (b) Competition variable $\Delta=x_{187}-x_{157}$: in the reference trajectory, $\Delta>0$, so that $x_{157}<x_{187}$ and site 157 is more unstable and is activated first. In the perturbed trajectory, $\Delta<0$, so that $x_{187}<x_{157}$ and site 187 is more unstable and is activated first. 
}
\label{fig5}
\end{figure}

The origin of the diffusive separation can be understood in the context of the post-yield diffusion and approximately Gaussian displacement statistics previously observed in these systems \cite{fiocco2013oscillatory,regev2019critical}. Here, however, the diffusing quantity is not the displacement of a single trajectory but the separation between two initially nearby trajectories. Before branching, the reference and perturbed trajectories follow the same sequence of plastic events and therefore remain strongly correlated.
Following a branching event, the trajectories follow different sequences of plastic events and decorrelate rapidly. Since the post-yield dynamics is itself diffusive, the separation between two decorrelated trajectories also exhibits diffusive growth. To verify that once branching occurs, the two trajectories are uncorrelated, we calculated the slope for the squared distance (Fig.~\ref{fig3}b) and verified that it is twice the diffusion coefficient of self-diffusion in the system (for more details see Supplementary Material).



As we have seen, the soft-spot model reproduces both the trajectory branching and the exponentially distributed waiting times observed in the particle-based simulations. The advantage of the model is that it allows us to follow the sequence of soft-spot activations directly and therefore identify the microscopic events associated with trajectory separation. 
To illustrate how branching is initiated, we examine the example shown in  Fig.~\ref{fig4}, which follows the early stages of a branching event occurring within an avalanche, during the increasing-strain branch of a forcing cycle with $\gamma_{\rm max}=1.2$.
The reference and perturbed trajectories visit the same sequence of configurations up to and including state $C$. At this point, however, different soft spots are activated. In the reference trajectory, soft spot 667 at site 157 is activated, whereas in the perturbed trajectory soft spot 863 at site 187 is activated. As a consequence, the two trajectories follow different sequences of plastic events and separate.
To understand why different soft spots are selected, we consider the distance to instability \cite{lin2014density,lin2016mean}:
\begin{equation}
x_i=\sigma_i^+-\sigma_i\,.
\end{equation}
Here $\sigma_i$ is the stress at site $i$ and $\sigma_i^+$ is the stress at which the soft spot in site $i$ becomes unstable. At each stable configuration, the soft spot with the smallest value of $x_i$ is the next to be activated. Inside an avalanche, the soft spot with the most negative $x_i$ is activated next by the dynamics.

Figure~\ref{fig5}a shows that, immediately before branching, sites 157 and 187 (highlighted by the arrows) are the only sites with negative values of $x_i$. In the reference trajectory, their values are $x_{157}=-2.915\times10^{-4}$ and $x_{187}=-2.913\times10^{-4}$, respectively, and differ by only $2\times10^{-7}$. Because this difference cannot be resolved on the scale of Fig.~\ref{fig5}a, we define the competition variable 
\begin{equation} 
\Delta=x_{187}-x_{157}. 
\end{equation}
Figure~\ref{fig5}b shows that in the reference trajectory $\Delta>0$, indicating that the soft spot at site 157 is more unstable and is therefore activated first. In the perturbed trajectory, however, $\Delta<0$, indicating that the soft spot at site 187 is activated first instead. Although the perturbation produces only very small changes in the individual distances to instability, it changes the outcome of the competition between the two nearly-degenerate instabilities and thereby alters the subsequent sequence of plastic events.
%

The example shown in Fig.~\ref{fig4} further demonstrates that the soft spot that loses the competition is not necessarily removed from the subsequent dynamics. In this realization, soft spot 863 eventually activates in both trajectories, but at different stages of the avalanche. We can therefore see that changes to timing and ordering of soft spot activations contribute to trajectory branching at the early stages of separation.

The competition mechanism and resulting changes in activation ordering identified here is closely related to the scrambling mechanism previously introduced in studies of memory in cyclically driven amorphous solids and other mechanical systems
\cite{szulc2022cooperative,bense2021complex,Lindeman2021Multiple}.
In those works, interactions between hysteretic elements led to changes in the order of activation events (``cascade-scrambling'' \cite{szulc2022cooperative}), which was identified as an important ingredient in the emergence of multi-periodic dynamics \cite{Keim2021Multiperiodic,szulc2022cooperative}. Here, a perturbation modifies the outcome of a competition between nearly-unstable instabilities, thereby scrambling the subsequent activation sequence and placing the system on a different trajectory. Unlike the multi-periodic case, where scrambling modifies the structure of recurrent dynamics, the resulting reordering here produces sustained trajectory separation.

Our results therefore suggest that branching originates from competition between unstable soft spots, whose outcome determines the subsequent ordering of plastic events. Small perturbations can alter the outcome of this competition and thereby scramble the ensuing activation sequence, providing a microscopic mechanism for random trajectory separation in a system that remains locally stable.

\section{Discussion}
We have shown that, unexpectedly, amorphous solids subjected to oscillatory shear above yielding do not exhibit exponential sensitivity to small perturbations. Instead, nearby trajectories initially contract and remain close for extended periods before abruptly separating by a rare branching event. Once branching occurs, the trajectories follow different sequences of plastic events and subsequently separate diffusively. These observations identify a route to irreversibility that does not rely on conventional chaotic divergence. 

We find that the waiting times between branching events are exponentially distributed and provide a minimal explanation for this behavior in terms of known statistical properties of plastic events in amorphous solids. We further show that a discrete, soft-spot-based model exhibits qualitatively similar behavior, indicating that the origin of the branching instability can be understood at the level of soft-spot dynamics. By directly following the evolution of individual soft spots during a branching event, we find that branching originates from a competition between nearly-degenerate plastic instabilities. A small perturbation to the stress field alters which of the competing instabilities activates first, thereby changing the ordering of subsequent plastic events. This places the perturbed system on a different trajectory and initiates sustained trajectory separation. We show that after branching, the trajectories become effectively uncorrelated, as demonstrated by the fact that the diffusion coefficient associated with trajectory separation equals twice the self-diffusion coefficient of a single trajectory.

Finally, we verified that limit cycles reached below yielding are stable to small perturbations, with no branching events observed. In contrast, the post-yield steady state exhibits a finite branching rate that increases with strain amplitude. This sharp distinction suggests that irreversibility arises when branching becomes a persistent feature of the dynamics. Whereas perturbations below yielding decay and the system returns to the same periodic trajectory, trajectories above yielding continue to branch at a finite rate, preventing convergence to a limit cycle.

These results bridge the gap between the stochastic description of the irreversibility transition and the underlying deterministic dynamics. While branching originates from a deterministic dynamical mechanism, namely competition between nearly-degenerate plastic instabilities, the disordered structure of the material generates a random distribution of instability thresholds. As a result, branching events occur at apparently random times and are well described by a stochastic rare-event process. The combination of deterministic instability selection and stochastic branching statistics therefore provides a natural connection between dynamical-systems and stochastic descriptions of the irreversibility transition.

Future work will focus on developing a statistical-mechanical description of the irreversibility transition in terms of branching dynamics and clarifying its connection to non equilibrium phase transitions \cite{di2017simple,grimmett2001probability,harris1963theory}. Another question concerns the role of overlaps between soft spots \cite{szulc2024overlapping} in sustaining irreversible dynamics. In the present work, we focus on the instability-selection mechanism that initiates trajectory branching. Determining whether and how soft-spot overlaps enable branching to produce persistent exploration of state space remains an important direction for future study.

\section{Acknowledgements}
\noindent
I.R. and S.C. were supported by the Israel Science Foundation (ISF) through Grant No. 1204/23.
I.R. would like to thank Thomas Witten, Haim Diamant, Golan Bel and Yosef Ashkenazy for useful discussions. 
\bibliographystyle{unsrt}
\bibliography{sens}
\section{Supplementary material}
\subsection{Simulation details}
\subsubsection{Particle based simulations}

We used LAMMPS \cite{thompson2022lammps} to simulate a system of $1000$ point particles in two dimensions, interacting via the Kob-Andersen $65:35$ Lennard Jones binary system \cite{bruning2009glass} (in LJ units):
\begin{equation}\label{eq:lj_potential}
U_{\alpha\beta}(r) = 4\epsilon_{\alpha\beta}\left[\left(\frac{\sigma_{\alpha\beta}}{r}\right)^{12} - \left(\frac{\sigma_{\alpha\beta}}{r}\right)^6\right]
\end{equation}
where $\alpha, \beta \in \{A, B\}$ denote the particle species. The interaction parameters are: $\sigma_{AA} = 1.0$, \ $\epsilon_{AA} = 1.0$ (reference scales) ; $\sigma_{AB} = 0.8$, \ $\epsilon_{AB} = 1.5$ (strong cross-interaction) ; $\sigma_{BB} = 0.88$, \ $\epsilon_{BB} = 0.5$ (small, weak B particles).  In all interactions, we applied a cutoff of 2.5. 
In all the simulations we used the same system size $L= 29.34$ (NVT ensemble). The initial amorphous samples were prepared by equilibrating a system of particles at a moderately high temperature $T=0.5$, and then instantaneously quenching to zero temperature using FIRE minimization algorithm. 
The resulting configurations were sheared quasistatically following the athermal quasi-static shear (AQS) protocol where small shear increments are applied using the Lees-Edwards periodic boundary conditions and small strain steps of $\delta\gamma = 10^{-4}$ are followed by FIRE energy minimization\cite{lees1972computer,FIRE}. In our simulations, we applied oscillatory forcing in the following way. Starting from zero strain, incremental positive strain steps are imposed until a predetermined maximum strain, $\gamma_{\rm max}$ is reached. The direction of strain is then reversed and incremental strain steps are applied in the opposite direction until $-\gamma_{\rm max}$ is reached. Finally, the strain is reversed once more to bring the system back to zero strain, thus completing one full cycle. This process is then repeated for subsequent cycles in the form: $0 \rightarrow \gamma_{\max} \rightarrow -\gamma_{\max} \rightarrow 0 \rightarrow \cdots$.

\subsubsection{Data preparation protocol}
\begin{figure}[htbp]
\centering
\includegraphics[width=\linewidth]{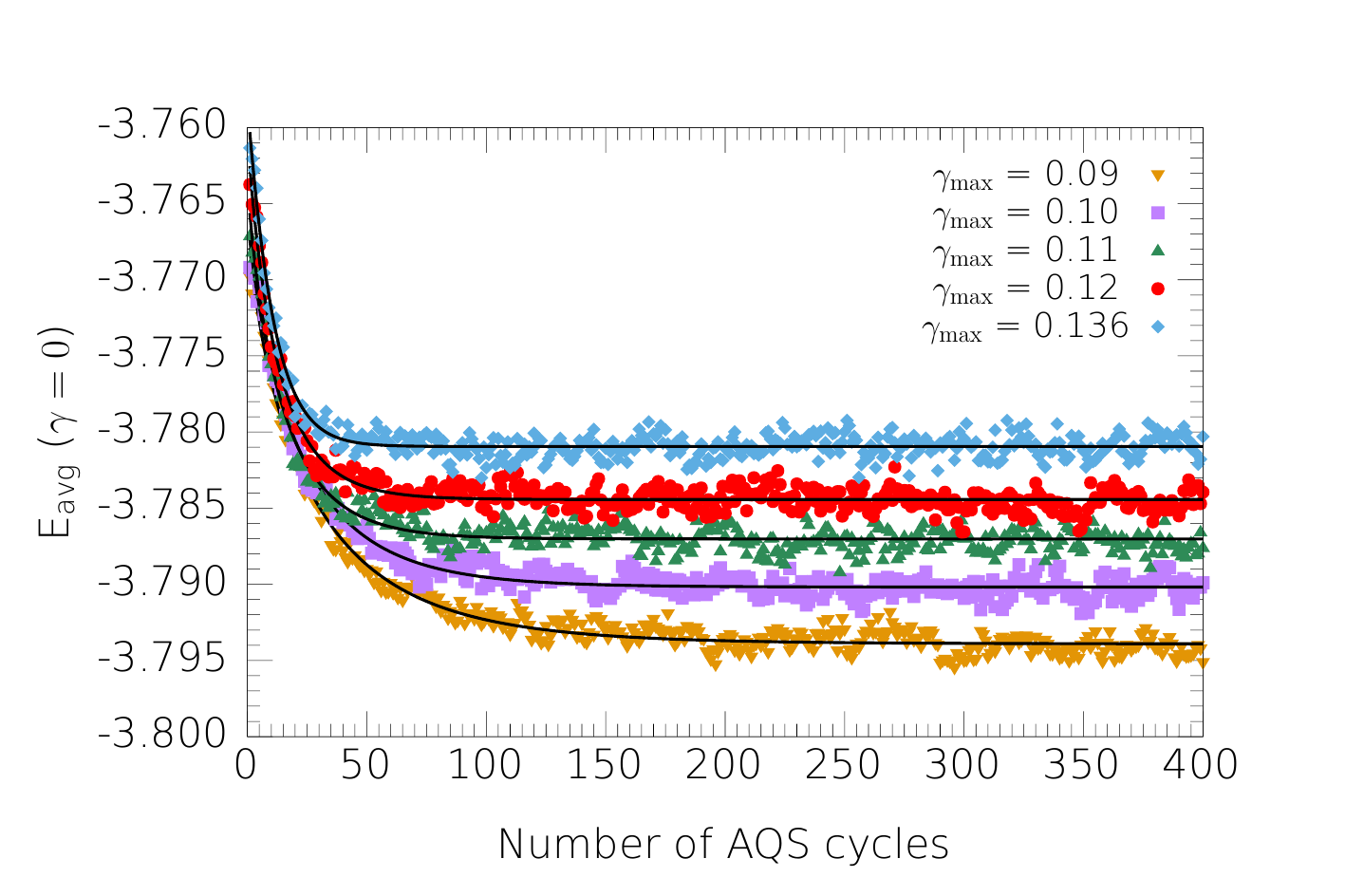}
\caption{
Evolution of the averaged potential energy per particle, $E$, measured at zero strain. Data are shown for varying strain amplitudes $\gamma_{\rm max}$: $0.09$ (yellow $\blacktriangledown$), $0.10$ (purple $\blacksquare$), $0.11$ (green $\blacktriangle$), $0.12$ (red $\bullet$), and $0.136$ (cyan $\blacklozenge$). Averages are taken over $100$ realizations. Solid lines indicate fits obtained using a stretched exponential model.
}
\label{avg_pe}
\end{figure}

As discussed in the text, the steady-state differs depending on whether $\gamma_{\rm max}$ is larger or lower than $\gamma_c$. In the sub-yield regime, the system settles into a reversible limit cycle, while in the post-yield regime, the system reaches an irreversible steady-state in which macroscopic quantities such as the potential energy, settle into a constant average value. Here, the steady state is attained by subjecting an ensemble of realizations to a large number of AQS cycles. In the post-yield regime, a steady state is identified as the point at which the averaged potential energy fluctuates around a constant plateau after an initial transient. For this study, we prepared $100$ realizations for each strain amplitude $\gamma_{\rm max} \in \{0.09, 0.10, 0.11, 0.12, 0.136\}$, with each realization subjected to $400$ AQS cycles. Although the transient duration (measured in cycles required to reach the constant plateau) decreases as $\gamma_{\rm max}$ increases, $400$ AQS cycles are sufficient to guarantee steady-state conditions for all the strain amplitudes, as demonstrated in Fig.~\ref{avg_pe}. 

%
%
\subsubsection{$\lambda_b$ calculation}
\begin{figure}[htbp]
\centering
\includegraphics[width=\linewidth]{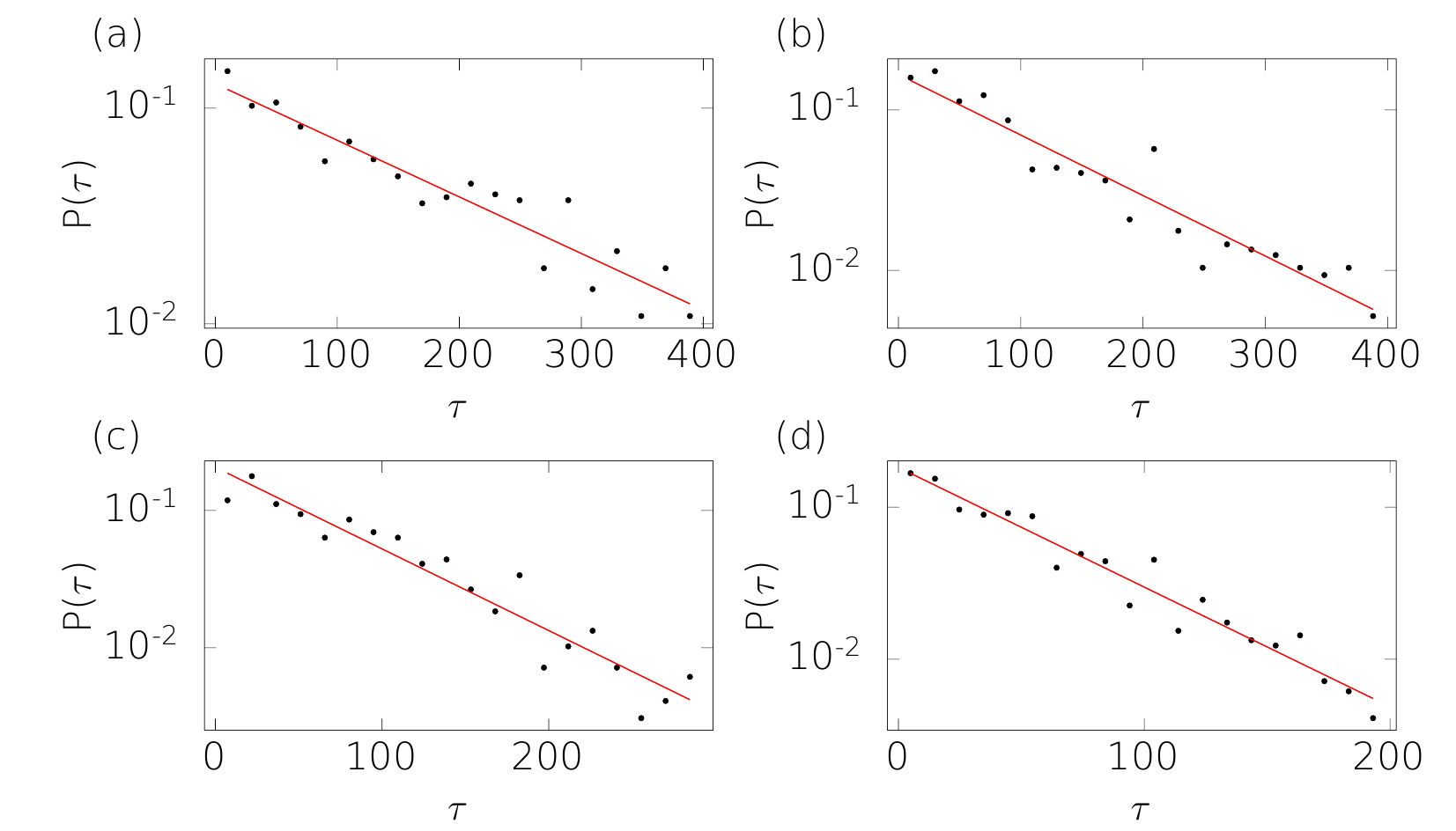}
\caption{
Distributions of the waiting time $\tau$ (measured in AQS cycles) for different, post-yield, strain amplitudes, obtained from particle-based simulations and averaged over $100$ realizations. The distributions display an exponential decay for (a) $\gamma_{\rm max} = 0.09$, (b) $0.10$, (c) $0.11$, and (d) $0.12$. (The data for $\gamma_{\rm max} = 0.136$ is presented in the main text).
}
\label{lambda_b}
\end{figure}

For $\gamma_{\rm max} = 0.136$ we found the waiting time distribution from 500 realizations running for 100 cycles. 
For the other post-yield amplitudes, where waiting times were larger, we found the waiting time distribution from 100 systems running for 400 forcing cycles. We observe that as the strain amplitude increases, the number of AQS cycles required for all the realizations to branch decreases. As can be seen in Fig.~\ref{lambda_b}, and discussed in the text, the characteristic branching rate, $\lambda_{b}$, increases gradually as $\gamma_{\rm max}$ increases. 

%
\subsubsection{Soft-spot model}
We use the hysteron-based model introduced in Ref.~\cite{szulc2024overlapping}. The elementary degrees of freedom are interacting, hysteretic two-state elements representing localized plastic rearrangements, or ``soft spots'', in amorphous solids \cite{szulc2022cooperative,Lindeman2021Multiple, Keim2021Multiperiodic,bense2021complex,van2021profusion, mungan2019networks}. In the following, we refer to these elements as soft spots. The system consists of $N=250$ sites, each carrying a local stress $\sigma_i$. Plastic activity is driven by changes in the local stress, and switching is stress activated. Under quasistatic loading, the strain is increased until one site becomes unstable and undergoes a plastic event. A strain increment $\delta\gamma$ causes a uniform stress increment $\delta\sigma$, with
\begin{equation} 
\delta\sigma=\mu\delta\gamma, 
\end{equation} 
where the shear modulus is set to $\mu=1$. Each site represents a group of soft spots that share some of their constituent particles. Consequently, the activation of one soft spot can block other soft spots at the same site through their mutual overlap. Each soft spot $n$ is a two-state element with state $s_n\in\{0,1\}$. It switches from $0$ to $1$ when the stress at its site reaches its upper threshold $\sigma_n^+$ and switches from $1$ to $0$ when the stress decreases to its lower threshold $\sigma_n^-$. In any configuration, each site has two switchable soft spots: one that can be activated under increasing stress and one that can be activated under decreasing stress. When the upward-switchable soft spot switches, it blocks its downward-switchable partner, and a new upward-switchable soft spot is created. Upon reversal of the drive, the most recently activated soft spot switches back, releases the blocked soft spot, and restores the previous configuration. This hierarchical, last-in-first-out organization is analogous to transitions between neighboring minima of a multi-well potential, as used in recent elastoplastic models \cite{kumar2022mapping,liu2022fate}. The switching thresholds of each soft spot are assigned upon its creation. They are drawn independently from a folded normal distribution, 
\begin{equation} 
\sigma_n^+ = 2|X^+|, \qquad \sigma_n^- = -2|X^-|, 
\end{equation} 
where 
\begin{equation} 
X^\pm\sim\mathcal{N}(0,0.09). 
\end{equation} 
Thus, $\sigma_n^+>0>\sigma_n^-$, and the local response is hysteretic by construction. When a soft spot reaches one of its switching thresholds, the stress at the corresponding site changes by 
\begin{equation} 
\sigma_i^{\rm drp} = 0.5\left(\sigma_n^+-\sigma_n^-\right)Y, \qquad Y\sim\mathcal{U}[0,1), 
\end{equation} 
where $\sigma_i^{\rm drp}$ is subtracted from $\sigma_i$ for a $0\to1$ transition and added to $\sigma_i$ for a $1\to0$ transition. The prefactor prevents loops in which the same soft spot switches back and forth indefinitely. The switching thresholds, interaction coefficients, and properties of newly generated soft spots are random and collectively define a disorder realization. Once this realization is fixed, the subsequent dynamics is deterministic: the evolution from a given system configuration under a specified driving protocol is uniquely determined. Randomness therefore enters through the construction of the disorder realization rather than through stochastic choices made during the subsequent dynamics. 

In the absence of competing soft spots, the hierarchical multi-well dynamics described above eventually reaches a limit cycle under cyclic driving at any driving amplitude \cite{szulc2024overlapping}. Atomistic simulations, however, reveal cases in which overlapping soft spots do not follow this hierarchical switching order \cite{szulc2024overlapping}. To represent such events, whenever the dynamics reaches a system-wide configuration that has not been visited previously, a competing soft spot $\tilde n$ is added with probability $p_{\rm ir}=0.01$. If a competing soft spot is added, its upper threshold is drawn such that %
\begin{equation} 
\sigma_{\tilde n}^{+}<\sigma_n^{+}, 
\end{equation} 
where $n$ denotes the incumbent soft spot. The competing soft spot therefore switches before the incumbent soft spot and blocks it, causing the site to follow an alternative branch in configuration space. The probabilistic addition of a competing soft spot is performed only on the first visit to a system-wide configuration. Once a configuration has been visited, its available soft spots and transition rules are retained, and no new competing soft spot is introduced upon subsequent visits; that is, we set $p_{\rm ir}=0$ for previously visited configurations. Thus, the randomly generated thresholds, interactions, and competing-soft-spot insertions define the disorder realization and the network of transitions as it is explored. For a fixed realization, revisiting the same configuration under the same driving direction produces the same transition, preserving the deterministic character of the dynamics.

The dynamics is event driven. Let $n_i$ denote the soft spot at site $i$ that is currently switchable in the direction of the imposed driving. At each stable configuration, we calculate the distance to the next instability, 
\begin{equation} 
x_i = \begin{cases} \sigma_{n_i}^+-\sigma_i, & s_{n_i}=0 \text{ and the strain is increasing}, \\[2mm] \sigma_i-\sigma_{n_i}^-, & s_{n_i}=1 \text{ and the strain is decreasing}. \end{cases} \nonumber
\end{equation} 
We identify the site with the smallest value of $x_i$, switch the corresponding soft spot, and advance the strain in the imposed direction by 
\begin{equation} 
\delta\gamma=\frac{\delta\sigma_i}{\mu}. 
\end{equation} 
Following a switching event at site $i$, the stress at every other site $j\neq i$ is updated according to 
\begin{equation} 
\sigma_j\rightarrow\sigma_j+G_{ji}, 
\end{equation} 
using the random frustrated mean-field interaction kernel 
\begin{equation} 
G_{ji} = \frac{\eta_{ji}}{\sqrt{N}} + \frac{\tilde{\eta}_j}{N}, \qquad \tilde{\eta}_j = -\sum_{i\neq j}\frac{\eta_{ji}}{\sqrt{N}}, 
\end{equation} 
where 
\begin{equation} 
\eta_{ji}\sim\mathcal{U}[-0.2,0.2).
\end{equation} 
The interaction coefficients are drawn as part of the disorder realization and remain fixed throughout the subsequent evolution. With this normalization, the residual sum vanishes in the large-$N$ limit, consistent with the stress-conserving character of the quadrupolar elastic fields generated by soft spots \cite{lin2014density,bocquet2009kinetic}. The resulting stress redistribution can destabilize additional sites and initiate an avalanche. After each switching event, we therefore identify the currently unstable soft spots through 
\begin{equation} 
|x_j|= \begin{cases} |\sigma_{n_j}^+-\sigma_j|, & s_{n_j}=0 \text{ and } \sigma_j>\sigma_{n_j}^+, \\[2mm] |\sigma_j-\sigma_{n_j}^-|, & s_{n_j}=1 \text{ and } \sigma_j<\sigma_{n_j}^- . 
\end{cases} 
\end{equation} 
Among the unstable soft spots, the one with the largest value of $|x_j|$ is activated next. Stress redistribution and switching are repeated until no unstable soft spots remain. As discussed in Ref.~\cite{szulc2024overlapping}, the model exhibits an irreversibility transition at the critical strain amplitude $\gamma_c\simeq 0.97$, where the time required to reach a limit cycle diverges. 

To probe trajectory stability, we apply small perturbations to the local stress field of an initial configuration. Specifically, 
\begin{equation} 
\sigma_i(0) \rightarrow \sigma_i(0) + \delta\frac{\xi_i}{\|\boldsymbol{\xi}\|_2}, \qquad \xi_i\sim\mathcal{N}(0,1), 
\end{equation} 
where $\delta=10^{-5}$ and \
\begin{equation} 
\|\boldsymbol{\xi}\|_2 = \left(\sum_{i=1}^{N}\xi_i^2\right)^{1/2}. 
\end{equation} 
The random direction $\boldsymbol{\xi}/\|\boldsymbol{\xi}\|_2$ is normalized so that every perturbed realization has the same perturbation magnitude, 
\begin{equation} 
\|\Delta\boldsymbol{\sigma}\|_2=\delta. 
\end{equation} 
The stress at an individual site is therefore shifted by a typical amount of order $\delta/\sqrt{N}$. 
The reference and perturbed copies share the same realization of the disorder, including the interaction coefficients, soft-spot thresholds, and random-number sequence used when previously unvisited configurations are encountered. Both copies evolve according to the same dynamical rules, so that the applied stress perturbation is the only difference between their initial states. Any subsequent trajectory separation therefore results from the dynamical response to that perturbation.
\subsection{Exponential decay}
\begin{figure}[htbp]
\centering
\includegraphics[width=\linewidth]{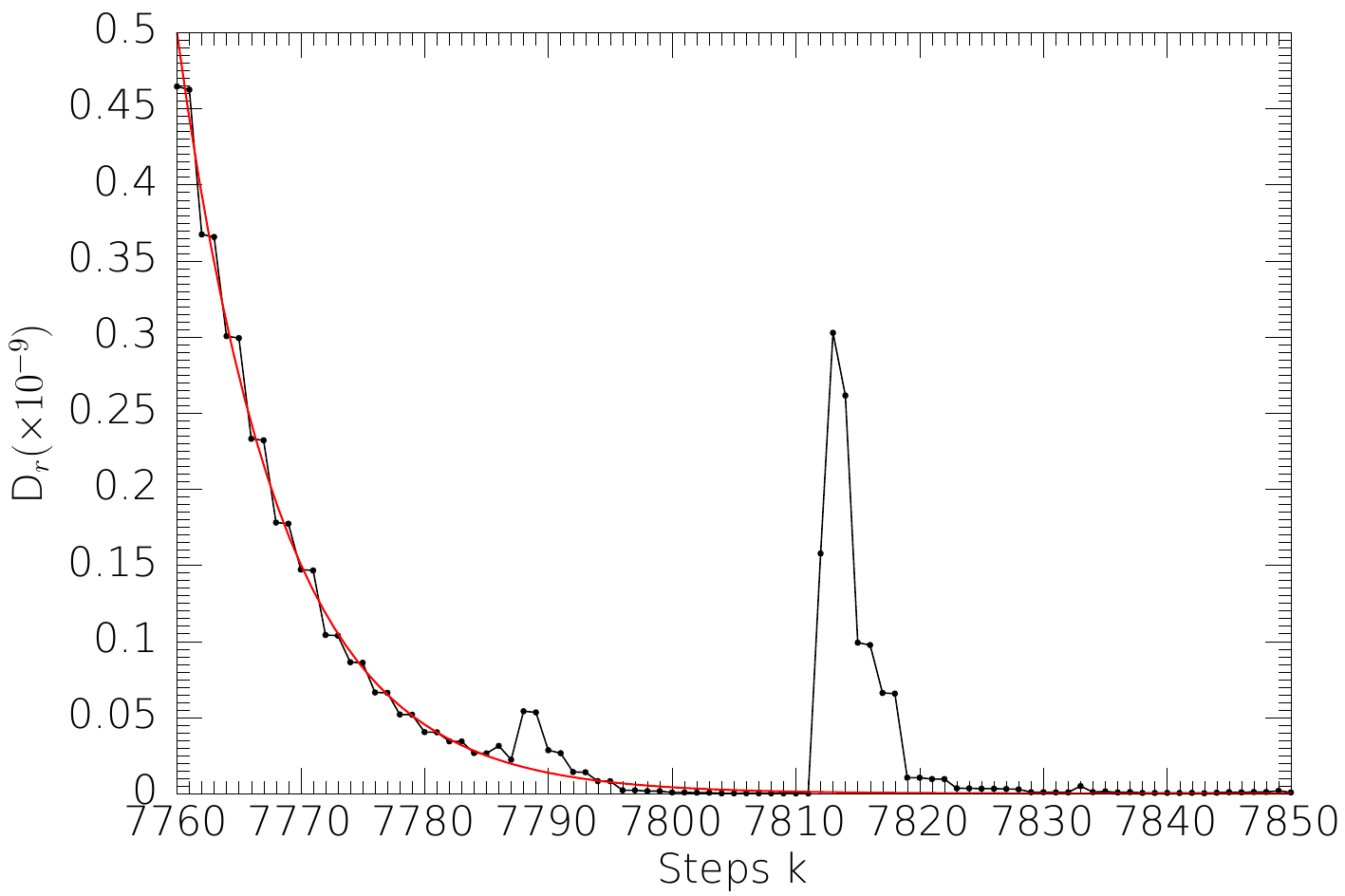}
\caption{
An exponential fit (red line) to a typical small, spontaneous deviation of a perturbed trajectory from the reference trajectory. 
}
\label{exponential}
\end{figure}

As discussed in the text, most deviations of a perturbed trajectory from a reference trajectory decay exponentially fast both in the sub-yield and post-yield regimes. In Fig.~\ref{exponential} we demonstrate it explicitly for a typical deviation in the post-yield regime.

%
\subsection{Theoretical derivations}
\subsubsection{Exponential waiting times}
Here we derive the exponential waiting-time distribution under the assumption that, in finite AQS systems, the strain intervals (and therefore the number of simulation steps) between successive plastic events are approximately exponentially distributed, consistent with the observations of Refs. \cite{maloney2006amorphous,tyukodi2019avalanches}. The time (or equivalently, strain) between plastic events it thus:
\begin{equation}
f(\Delta t) = \lambda_{\rm pl}e^{-\lambda_{\rm pl}\Delta t}.
\end{equation}
Where $\Delta t$ is the time (number of simulation steps) between plastic events and $\lambda_{\text{pl}}$ is the rate at which plastic events occur.
The time to get $\mathcal{N}$ plastic events is:
\begin{equation}
T_n = \sum_{i=1}^\mathcal{N} \Delta t_i
\end{equation}
and its distribution is the Erlang distribution \cite{feller1958introduction}:
\begin{equation} 
f_{T\mid\mathcal{N}=n}(t) = \frac{ \lambda_{\rm pl}^{n}t^{n-1} e^{-\lambda_{\rm pl}t} }{ (n-1)! }, \qquad n=1,2,\ldots . 
\end{equation}
Assuming that each plastic event independently initiates branching with a constant probability $p_b$, the number of plastic events up to and including the first branching event is geometrically distributed, 
\begin{equation} 
\Pr(\mathcal{N}=n) = p_b(1-p_b)^{n-1}, \qquad n=1,2,\ldots . 
\end{equation}
%
The distribution of $T$ is then a compound distribution:
\begin{align} f_T(t) &= \sum_{n=1}^{\infty} \Pr(\mathcal{N}=n) f_{T\mid\mathcal{N}=n}(t) \nonumber\\ &= \sum_{n=1}^{\infty} p_b(1-p_b)^{n-1} \frac{ \lambda_{\rm pl}^{n}t^{n-1} e^{-\lambda_{\rm pl}t} }{ (n-1)! } \nonumber\\ &= p_b\lambda_{\rm pl}e^{-\lambda_{\rm pl}t} \sum_{n=1}^{\infty} \frac{ \left[(1-p_b)\lambda_{\rm pl}t\right]^{n-1} }{ (n-1)! } \nonumber\\ &= p_b\lambda_{\rm pl}e^{-\lambda_{\rm pl}t} e^{(1-p_b)\lambda_{\rm pl}t} \nonumber\\ &= p_b\lambda_{\rm pl}e^{-p_b\lambda_{\rm pl}t}. 
\end{align}
%
%
Therefore, the waiting time $T$ follows an exponential distribution with rate $p_b\lambda_{\text{pl}}$:
\begin{equation}
f_T(t) = p_b\lambda_{\text{pl}} e^{-p_b\lambda_{\text{pl}} t}
\end{equation}
which is consistent with our observations (Fig.~\ref{fig2}b,d). This is a known compound distribution as discussed in Feller \cite{feller1958introduction} and other probability texts.
%
\begin{figure}[htbp]
\centering
\includegraphics[width=\linewidth]{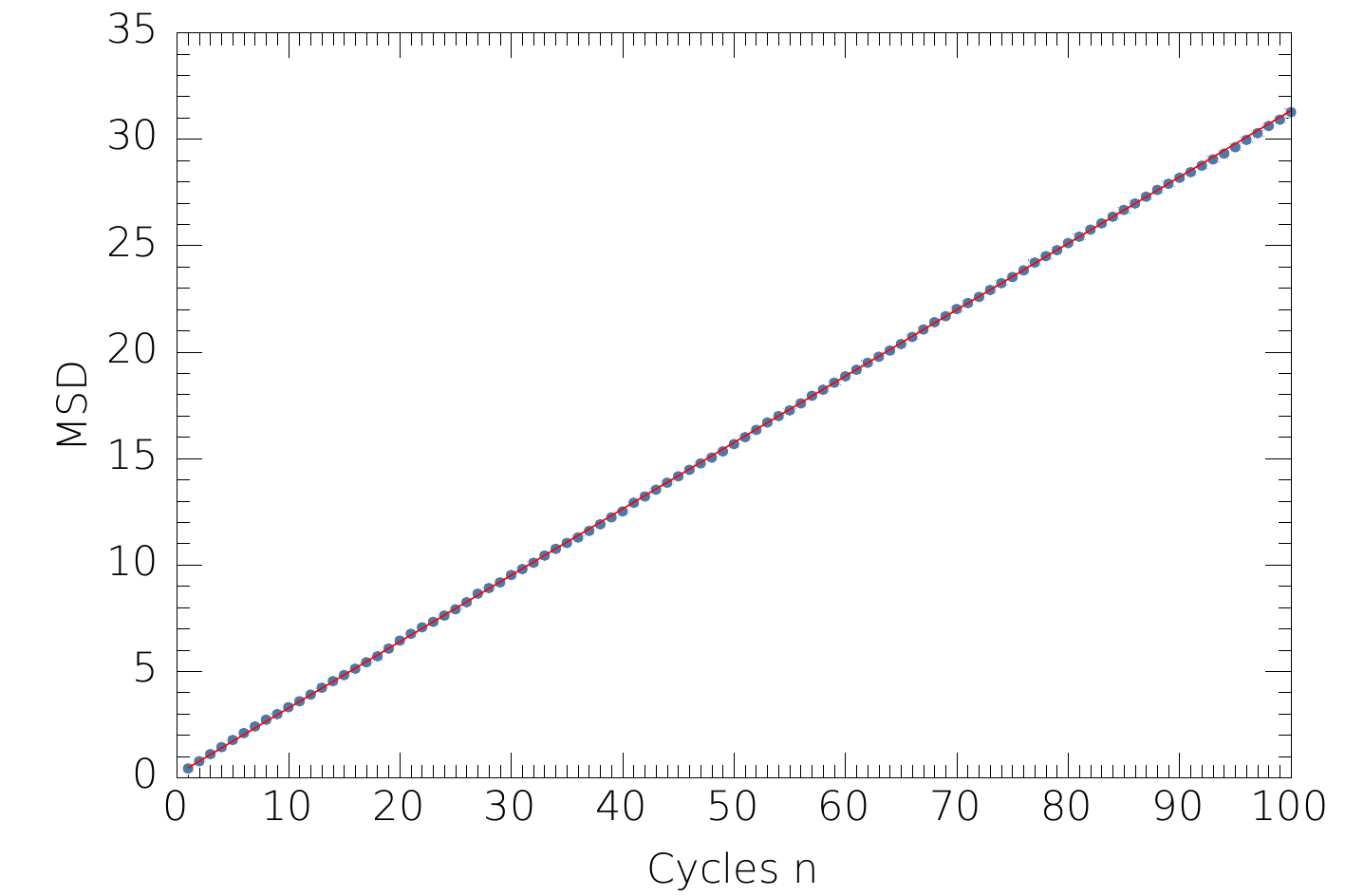}
\caption{
The MSD as measured from 100 realizations and 1000 perturbed trajectories. Fitting this curve, we obtain the diffusion coefficient $4D = 0.3119 \pm 0.0003$. 
}
\label{MSD}
\end{figure}
\subsubsection{Diffusive separation}
Previous studies have shown that particle motion in the post-yield regime is diffusive and exhibits Gaussian displacement statistics \cite{fiocco2013oscillatory,regev2019critical}. Here, however, the diffusing quantity is not the displacement of a single trajectory but the separation between two initially nearby trajectories.
Before branching, the reference and perturbed trajectories follow the same sequence of plastic events and therefore remain strongly correlated. After branching, the two trajectories follow different sequences of plastic events and accumulate different displacements. Let
\begin{equation}
\Delta {\bf R}_n
=
{\bf R}_n^{(p)}
-
{\bf R}_n^{(r)}
\end{equation}
denote the separation between the perturbed and reference trajectories after $n$ cycles. This separation can be expressed as
\begin{equation}
\Delta {\bf R}_n
=
\sum_{i=1}^{n}
\left(
\Delta {\bf r}_i^{(p)}
-
\Delta {\bf r}_i^{(r)}
\right),
\end{equation}
where $\Delta {\bf r}_i^{(p)}$ and $\Delta {\bf r}_i^{(r)}$ are the displacements accumulated during cycle $i$ in the perturbed and reference trajectories, respectively.
If branching causes the subsequent plastic-event sequences to become approximately decorrelated, then the displacement differences satisfy
\begin{equation}
\left\langle
\left(
\Delta {\bf r}_i^{(p)}
-
\Delta {\bf r}_i^{(r)}
\right)
\cdot
\left(
\Delta {\bf r}_j^{(p)}
-
\Delta {\bf r}_j^{(r)}
\right)
\right\rangle
\end{equation}
\begin{equation}
= \langle\Delta {\bf r}_i^{(p)}\cdot\Delta {\bf r}_j^{(p)}\rangle +  \langle\Delta {\bf r}_i^{(r)}\cdot\Delta {\bf r}_j^{(r)}\rangle = 
\end{equation}
\begin{equation}
= 4D\delta_{ij}  + 4D\delta_{ij} =  8D\delta_{ij}.
\end{equation}\\
in two dimensions, where $D$ is the self-diffusion coefficient. 
It then follows that
\begin{equation}
\left\langle
\Delta {\bf R}_n^2
\right\rangle
= 8D n.
\end{equation}
Thus, we expect:
\begin{equation}
\mathcal{D}^2_r(n) = 
\left\langle
\Delta {\bf R}_n^2
\right\rangle
= S n = 8D n\,
\end{equation}
where $S$ is a constant. This relation between $D^2_r(n)$ and $n$ is consistent with the diffusive scaling observed in the simulations. In this picture, diffusion originates from the progressive decorrelation of the plastic-event sequences followed by the two trajectories after a branching event. To verify that $S = 8D$, we calculated the diffusion coefficient from the mean-square displacement:
\begin{equation}
\langle {\bf r}^2_n\rangle = \langle \frac{1}{\mathcal{N}}\sum_{i=1}^\mathcal{N} |{\bf r}_i(n) - {\bf r}_i(0)|^2\rangle = 4Dn
\end{equation}
where ${\bf r}_i = (x_i,y_i)$.
The MSD and a linear fit are shown in Fig.~\ref{MSD}. The slope is 4 times the diffusion coefficient, which comes up to be $4D = 0.3119 \pm 0.0003$. From fitting the square distance (Fig.~\ref{fig3}b), we found that its slope is $S= 0.627 \pm 0.002$. 
Since $S/4D\approx 2.01$ we conclude that to a good level of confidence, $S = 8D$, confirming that post-branching trajectories are diffusive and statistically independent. 
\end{document}